\documentclass[traditabstract]{aa}
\usepackage{txfonts,graphicx,epsfig,times,natbib,array}
\bibpunct{(}{)}{;}{a}{}{,}

 \newcommand{\chandra}{{\it
    Chandra} } \newcommand{\asca}{{\it ASCA} }  \newcommand{\rosat}{{\it ROSAT} }
\newcommand{\xmm}{{\it XMM-Newton} }

\title{Metallicity map of the galaxy cluster A3667}
\titlerunning{Metallicity map of the galaxy cluster A3667}

\author{L. Lovisari\inst{1} \and W. Kapferer\inst{1} \and
  S. Schindler\inst{1} \and C. Ferrari\inst{2}} \authorrunning{L.
  Lovisari et al.}

\institute{
  Institut f{\"u}r Astro- und Teilchenphysik,
  Universit{\"a}t Innsbruck, Technikerstr. 25, A-6020 Innsbruck,
  Austria \and Universit\'e de Nice Sophia Antipolis, CNRS, Observatoire de la C\^ote
d'Azur, B.P. 4229, 06304 Nice Cedex 4, France}

\offprints{L. Lovisari}

\mail{Lorenzo.Lovisari@uibk.ac.at}

\date{Submitted , accepted }

\begin{document}

\abstract{We use \xmm data of the merging cluster Abell 3667 to
  analyze its metallicity distribution. A detailed abundance map of
  the central 1.1$\times$1.1 Mpc region indicates that metals are
  inhomogeneously distributed in the cluster showing a non-uniform and
  very complex metal pattern. The highest peak in the map corresponds
  to a cold region, slightly offset South of the X-ray center. This
  could be interpreted as stripped gas due to a merger between a group
  moving from NW towards the SE and the main cluster. We note several
  clumps of high metallicity also in the opposite direction with
  respect to the X-ray peak. Furthermore we determined abundances for
  5 elements (O, Si, S, Ar, Fe) in four different regions of the
  cluster. Comparisons between these observed abundances and
  theoretical supernovae yields allow to get constraints on the
  relative number of SN Ia and II contributing to the enrichment of
  the intra-cluster medium. To reproduce the observed abundances of
  the best determined elements (Fe, O and Si) in a region of
  7$^{\prime}$ around the X-ray center, 65-80$\%$ of SN II are
  needed. The comparison between the metal map, a galaxy density
    map obtained using 550 spectroscopically confirmed cluster members
    and our simulations suggest a recent merger between the main
    cluster and the group in the SE.

  }

\maketitle


\section{Introduction}
Cluster of galaxies are the largest virialized objects in the
universe. They form via gravitational instability from the initial
perturbations in the matter density field. The cosmic baryons fall
into the gravitational potential of the cluster dark matter halo
formed in this way, while the collapse heat up the intra-cluster
medium (ICM). At the high temperature measured in rich cluster, kT$>$3
keV, the ICM is highly ionised and its spectrum presents several
emission lines, among which the most prominent is the Fe K-shell line
at $\sim$ 7 keV. As heavy elements are only produced in stars the
processed material must have been ejected into the ICM by cluster
galaxies. \\ There is more and more observational evidence that
various types of processes are at work (e.g. ram pressure stripping
and galactic winds; see \citealt{2008SSRv..134..363S} for a review),
which remove interstellar medium (ISM) from the galaxies. Hence, the
study of the metal distribution is a sensible way to better
understand the thermodynamical properties of the diffuse gas and the
past history of star formation in galaxy clusters
(\citealt{1992A&A...254...49A}; \citealt{1993ApJ...419...52R};
\citealt{2004cgpc.symp..260R}). \\ Simulations have shown that the
metal distribution in a cluster shows many stripes and blobs at the
positions where the enrichment has taken place. Depending on the
dynamical state of the cluster the metal distribution looks very
different: in simple merger configuration (e.g. collision between two
subclusters) pre-mergers have a metallicity gap between the
subclusters, post-mergers have a high metallicity between subclusters
(\citealt{2006A&A...447..827K}; \citealt{2007A&A...466..813K};
\citealt{2008ChJAS...8...93S}; \citealt{2008SSRv..134..363S}). \\
Although to obtain metal maps from observation is not easy because
a lot of photons have to be accumulated in each region to measure the
metal abundance, several groups, using \xmm and \chandra data, have
derived quite detailed metallicity maps
(\citealt{2002MNRAS.337...71S}; \citealt{2004MNRAS.349..952S};
\citealt{2005A&A...432..809D}; \citealt{2005MNRAS.357.1134O};
\citealt{2005A&A...444..673S}; \citealt{2006A&A...449..475W};
\citealt{2006MNRAS.371.1483S}; \citealt{2006PASJ...58..695H};
\citealt{2009A&A...493..409S}). In general these 2D maps show that the
distribution of metals in cluster is not spherically symmetric, but it
has several maxima and complex metal patterns. The range of
metallicities measured in a cluster from minimum to maximum comprises
easily a factor of two. \\ Because clusters retain all the metals
provided by their galaxies, X-ray measurements of the abundances of
each element in the ICM enable us to examine the ratio of SN Ia and
II. \\ In this paper we present the results of the analysis of the
cluster Abell 3667 observed with \xmm. The main aim of the work is to
study its metal distribution, that can reveal clues about the
chemical enrichment history of the cluster
(\citealt{2008ChJAS...8...93S}; \citealt{2008SSRv..134..379B}), and to
infer its dynamical state by comparison with previous results and
hydrodynamic simulations (\citealt{2006A&A...447..827K};
\citealt{2007A&A...466..813K}). Secondly, we determine the abundances
of several elements and using the yields of Supernovae type Ia and II
we try to understand the origin of metals in galaxy clusters. \\ The
paper is structured as follows: in Sect. 2 we give an overview of
A3667; in Sect. 3 we present the data sets and data reduction
techniques employed and we describe the X-ray image; in Sect. 4 we
present the metallicity map and relative comparison with simulations;
in Sect 5 we present measurements of metal abundances and SN ratio
determination, and in Sect 6 we discuss our results. A summary of our
conclusion is given in Sect. 7. \\ Throughout the paper we assume
H$_{0}$=70 km s$^{-1}$ Mpc$^{-1}$, $\Omega_{\Lambda}$=0.73 and
$\Omega_{M}$=0.27. At the nominal redshift of A3667 (z=0.055), the
luminosity distance is 245.7 Mpc and the angular scale is 64 kpc per
arcmin.

\begin{figure}
\epsfig{figure=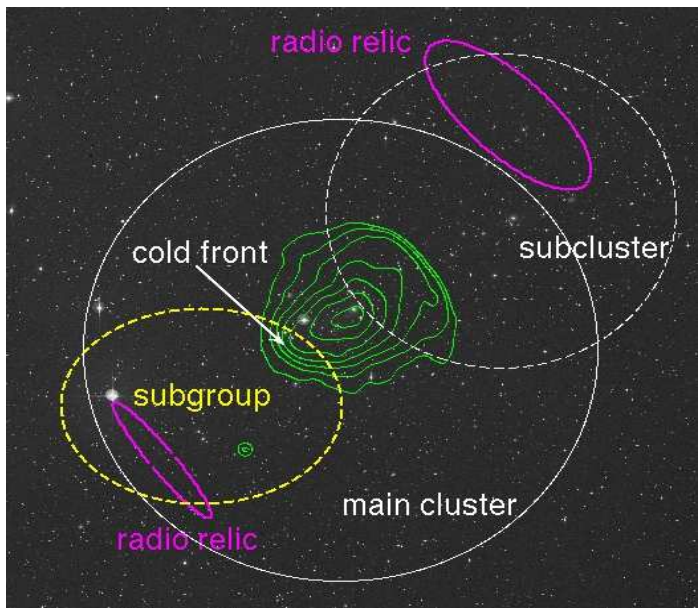,width=0.5\textwidth,height=8cm}
\caption{X-ray contours ($green$) overlaid on the optical Digital Sky Survey
  image. The circles represent the main cluster ($white \ full \
  line$) and the northwest subcluster ($dashed \ line$); the ellipses
  represent the subgroup of 27 members ($dashed \ yellow \ line$)
  found by \citet{2009ApJ...693..901O} and the two radio relics
  ($magenta \ lines$). The image is 1$^{\circ}\times1^{\circ}$.}
\label{fig:dss.eps}
\end{figure}

\section{Abell 3667}
A3667, a rich southern cluster at z=0.055 \citep{1992MNRAS.259..233S},
is famous for the presence of two extended radio relics symmetrically
located in the cluster periphery in the direction of the elongated
X-ray axis \citep{1997MNRAS.290..577R}. It has a large velocity
dispersion (\citealt{1988A&AS...72..415P},
\citealt{1992MNRAS.259..233S}, \citealt{1998ApJ...505...74G}) and the
2D galaxy distribution is bimodal (\citealt{1988A&AS...72..415P},
\citealt{1992MNRAS.259..233S}, \citealt{2008A&A...479....1J}), with
the main component around the cD galaxy near the X-ray peak and the
secondary component around the second brightest galaxy located in the
northwest, $\sim$15$^{\prime}$ from the X-ray
center. \citet{2009ApJ...693..901O}, with a large number of confirmed
galaxy members (550), found evidence for a new subgroup located in the
southeast region (see Fig. \ref{fig:dss.eps}). The bimodal structure
is evident also in the weak lensing mass map
\citep{2000ApJ...534L.131J} where it is possible to see a significant
mass concentration in the southeast of the cluster but not coincident
with the substructure shown by \citet{2009ApJ...693..901O}. \rosat and
\asca observations have shown a distorted X-ray morphology in the direction
of the reported bimodal optical distribution
(\citealt{1996ApJ...472..125K}, \citealt{1999ApJ...521..526M}). \xmm
and \chandra observations have revealed an inhomogeneous temperature
structure (\citealt{2002ApJ...569L..31M},
\citealt{2004A&A...426....1B}) and evidence for a cold front
\citep{2001ApJ...551..160V}. All these features indicate that the
cluster suffered a merger recently. Two different scenarios were
suggested by \citet{2009ApJ...693..901O}. The first is a two-body
merger between similar mass structures (the ``main cluster'' and the
NW ``subcluster'' in Fig. \ref{fig:dss.eps}) taking place in the plane
of the sky. In this scenario the NW subcluster would have already
traversed the main cluster along a SE-NW direction, producing two
outgoing shocks that would account for the double radio relics
observed in A3667. In this case, both the cold front and the subgroup
in the southeast of Fig. \ref{fig:dss.eps} could have been associated
with the northwest cluster, and then sloshed out or tidally stripped
during the passage through the core of the main cluster. The SE
subgroup could also be a background or foreground structure. An
alternative scenario involves a three-body merger between the main
cluster and the NW and SE substructures along a NW-SE axis. The NW and
SE radio relics would then be associated to the merger between the
main cluster and the NW and SE subclusters respectively, and the cold
front could be the remnant cold core of the SE subgroup.

\begin{figure}
  \epsfig{figure=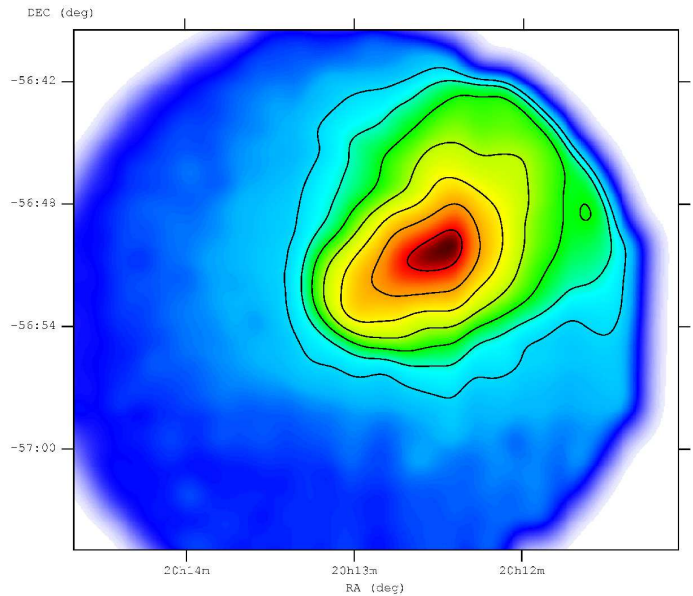,width=0.5\textwidth,height=8cm}
  \caption{Total (MOS+pn) EPIC mosaic image of A3667 in the [0.5-8]
    keV. The image is non-background subtracted, corrected for
    vignetting and exposure and adaptively smoothed. Contours indicate
    the surface brightness spaced by a factor of $\sqrt{2}$ starting
    at 5.5$\times$10$^{-5}$ cnts s$^{-1}$ arcmin$^{-2}$ level, which
    is 3.7 times the average background level. North is up and east is
    to the left, as they are in all other images. Coordinate grids are
    shown for the
    J2000. } \label{figure:a3667smoothed.ps} \end{figure}

\section{Observations and data reduction}
\subsection{X-ray analysis}
Observation data files (ODFs) were retrieved from the XMM archive and
reprocessed with the \xmm Science Analysis System (SAS) v7.1.0. We
used tasks $emchain$ and $epchain$ to generate calibrated event files
from raw data. Throughout this analysis single pixel events for the pn
data (PATTERN 0) are selected, while for the MOS data sets the
PATTERNs 0-12 are used. In addition, for all cameras events next to
CCD edges and next to bad pixels were excluded (FLAG==0). \\ The data
were cleaned for periods of high background due to the soft proton
solar flares using a two stage filtering process. We first accumulated
in 100 s bins the light curve in the [10-12] keV band for MOS and
[12-14] keV for pn, where the emission is dominated by the
particle-induced background, and exclude all the intervals of exposure
time having a count-rate higher than a certain threshold value (the
chosen threshold values are 0.20 cps for MOS and 0.25 cps for pn).
After filtering using the good time intervals from this screening, the
event lists was then re-filtered in a second pass as a safety check
for possible flares with soft spectra (\citealt{2005ApJ...629..172N};
\citealt{2005A&A...443..721P}). In this case light curves were made
with 10 s bins in the full [0.3-10] keV band. The resulting exposure
times after cleaning are 56.6 ks for MOS1, 56.1 ks for MOS2 and 45.7
ks for pn. \\ To correct for the vignetting effect, we used the photon
weighting method \citep{2001A&A...365L..80A}. The weight coefficients
were computing by applying the SAS task $evigweight$ to each event
file. Point sources were detected using the task $ewavelet$ in the
energy band [0.3-10] keV and checked by eye on images generated for
each detector. We produced a list of selected point sources from all
available detectors and the events in the corresponding regions were
removed from the event lists. \\ The background estimates were
obtained using the dedicated blank-sky event lists accumulated by
\citet{2003A&A...409..395R}. The blank-sky background events were
selected by applying the same PATTERN selection, vignetting
correction, flare rejection criteria and point source removal used for
the observation events. In addition, we transformed the coordinates of
the background file such that they were the same as for the associated
cluster data set. The background subtraction was performed using the
double subtraction process described in full detail in
\citet{2002A&A...390...27A}. It involves subtraction of the normalised
blank field data, and subsequent subtraction of the cosmic X-ray
background component estimated in the area of the field of view that
does not show cluster emission.

\subsection{X-ray image}
As the X-ray morphology can give interesting qualitative (and
quantitative, see e.g., \citealt{1996ApJ...458...27B}) insights into
the dynamical status of a given cluster, the adaptively smoothed,
exposure corrected (MOS+pn) count rate image in the [0.5-8] keV energy
band is presented in Fig. \ref{figure:a3667smoothed.ps}. The smoothed
images was obtained from the raw image corrected for the exposure map
by running the task $asmooth$ set to desired signal-to-noise ratio of
20. Regions exposed with less than 10$\%$ of the total exposure were
not considered. It is possible to see the elongated structure of the
cluster and the cold front to the southeast extensively discussed by
\citet{2001ApJ...551..160V}. The distortion in the northwest is
introduced by the field of view (FOV) edge. In Fig. \ref{fig:dss.eps}
we show the X-ray contours superposed on the optical image of the
cluster. The X-ray peak lies at RA 20:12.:27 and DEC -56:50:11 (J2000)
and is near the central dominant cluster galaxy located at RA 20:12:27
and DEC -56:49:36 (\citealt{2007MNRAS.381..494O}).

\begin{figure}
  \epsfig{figure=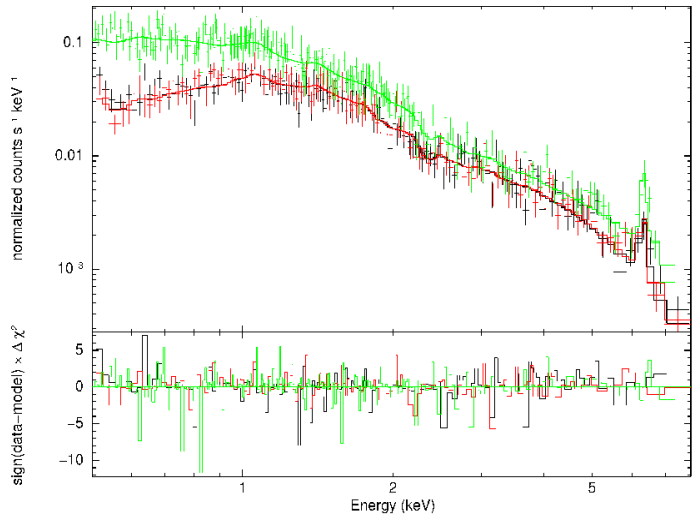,width=0.5\textwidth,height=7cm}
  \caption{Example of one EPIC pn ($green$) MOS1 ($black$) and MOS2
   ($red$) spectrum used to determine the abundance for
   metallicity map with an error of 10-20$\%$. } \label{fig:57-57.ps}
\end{figure}

\section{Metallicity map}
\begin{figure}
  \vbox{
    \epsfig{figure=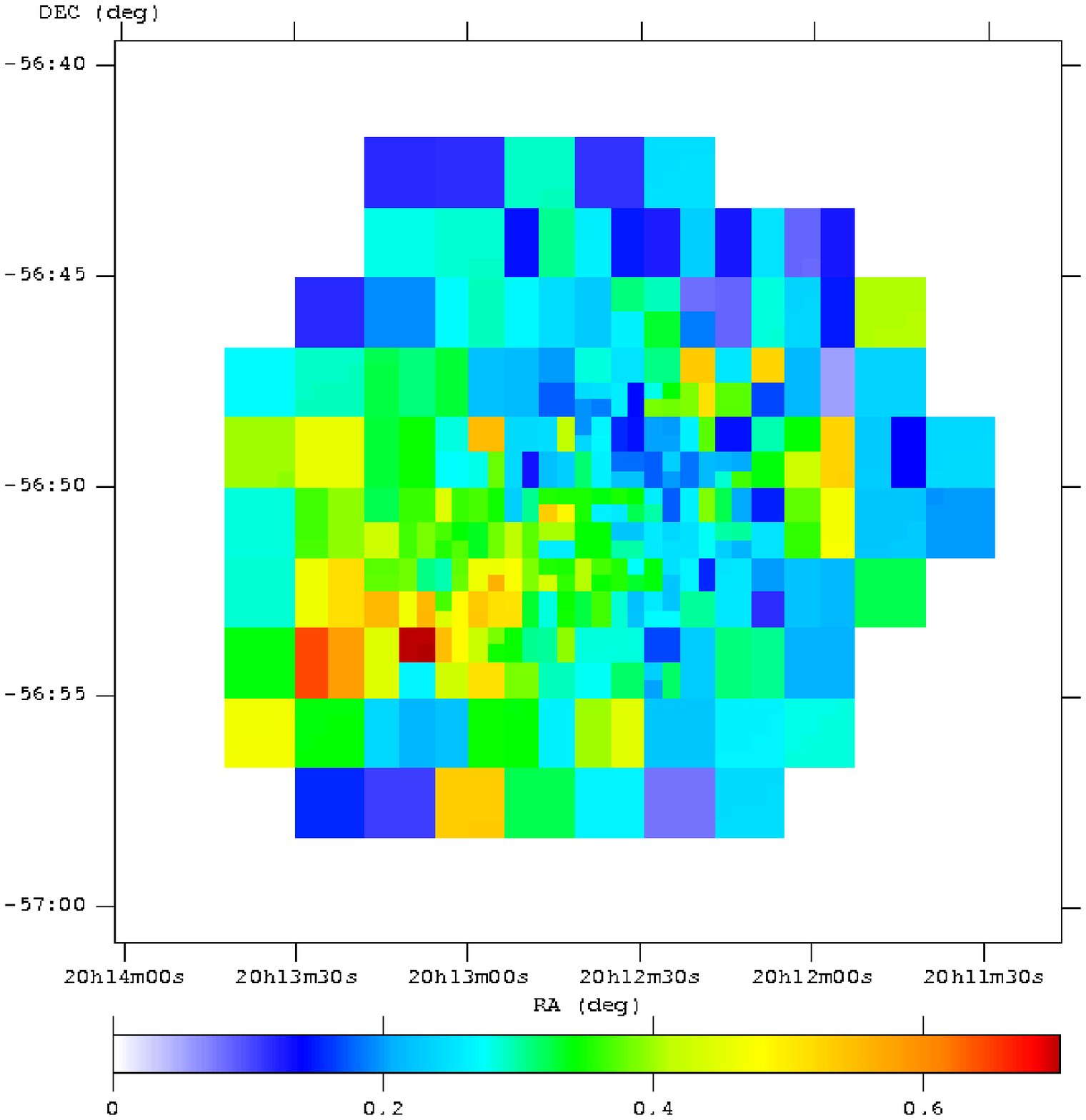,width=0.5\textwidth,height=8.5cm}
 \epsfig{figure=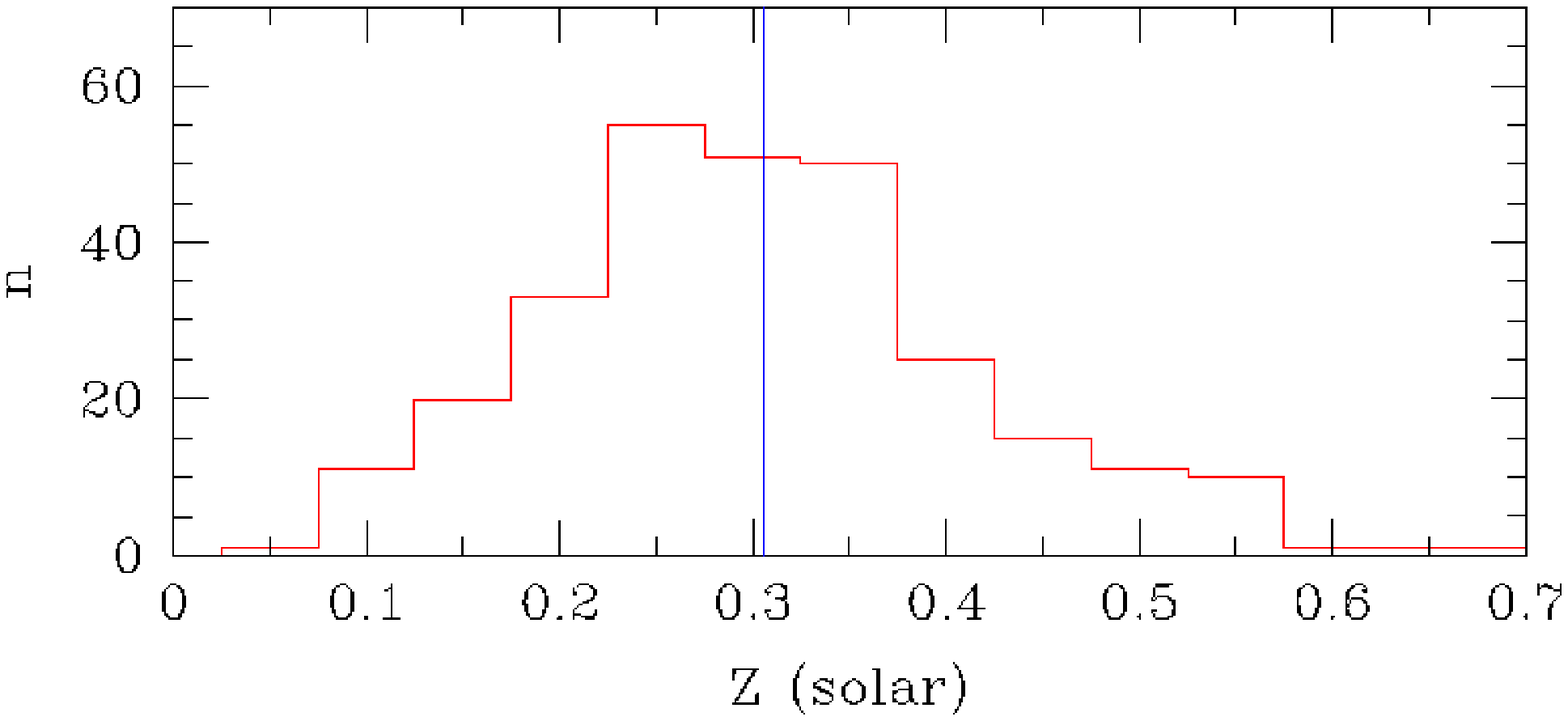,width=0.5\textwidth}
}
\caption{$Upper \ panel$: Metallicity map based on spectra from all
  three EPIC camera. The scale for the metallicity is
  in solar units. $Lower \ panel$: Number of bins with a certain
  metallicity. The vertical line in blue represents the mean value. }
\label{fig:zzend.ps} \end{figure}

\begin{figure}
\begin{center}
  \vbox{
    \epsfig{figure=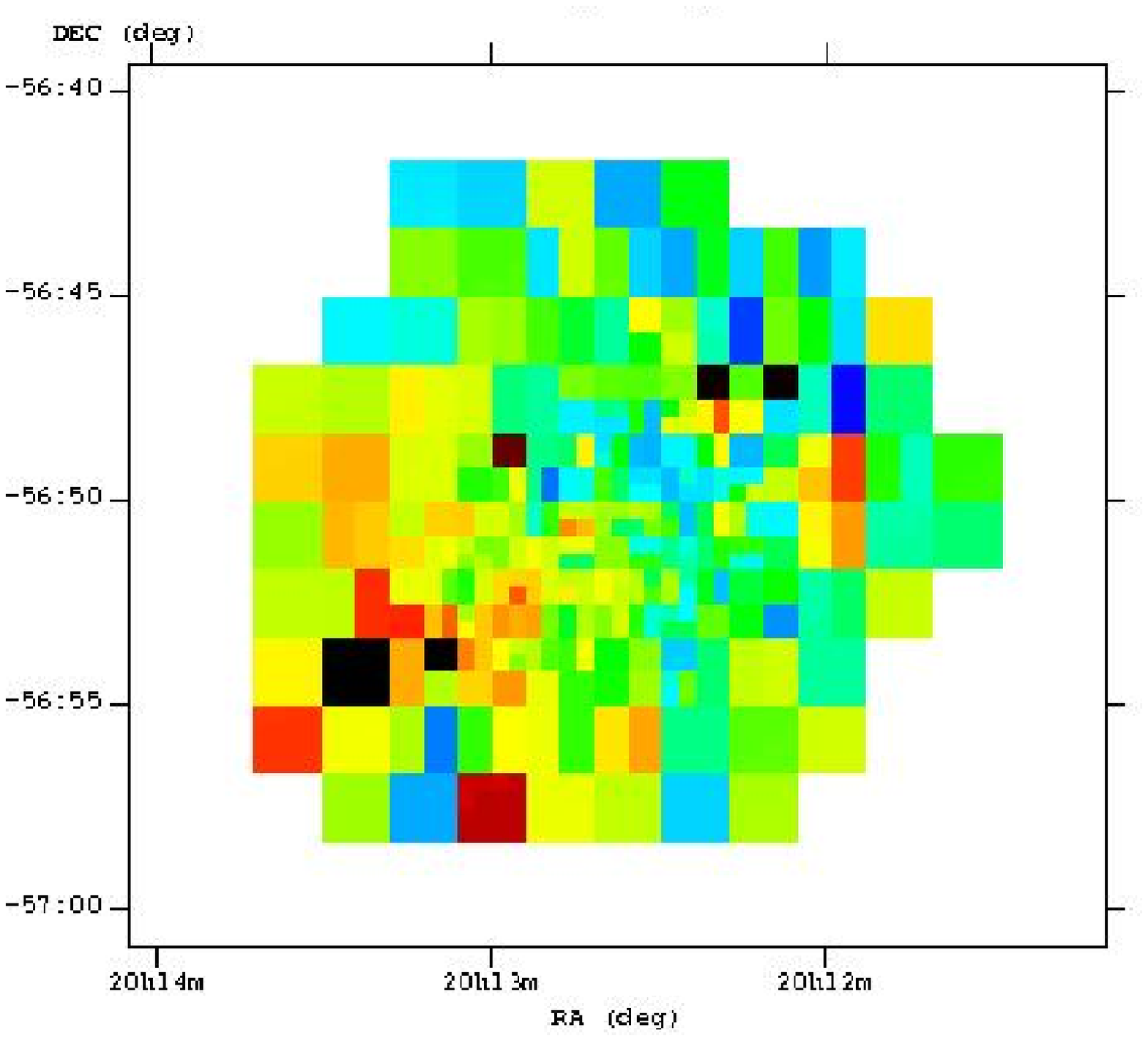,width=8.5cm,height=7cm}
    \epsfig{figure=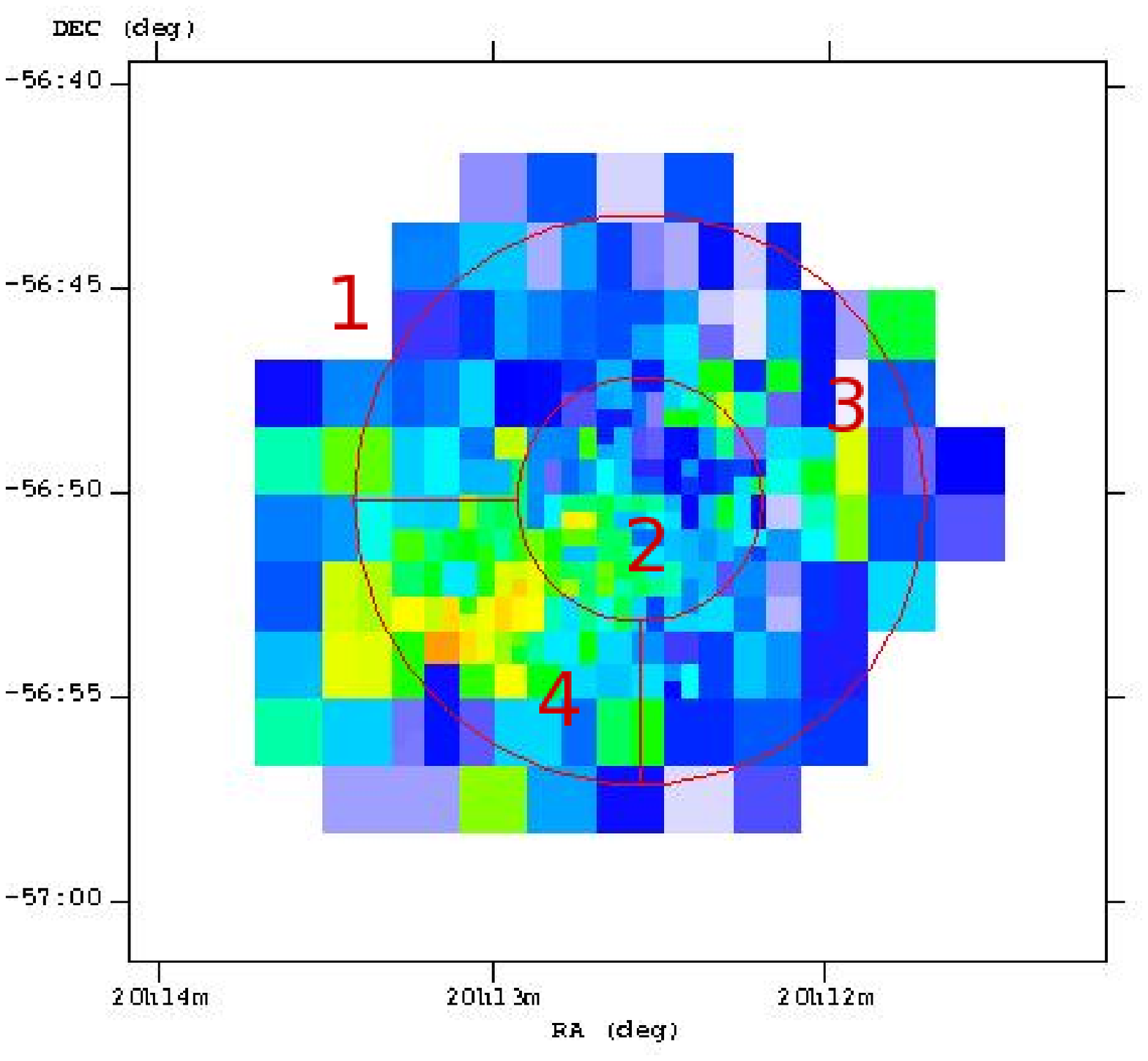,width=8.5cm,height=7cm} }
  \caption{Metallicity map for 1$\sigma$ upper ($top$) and lower
    ($bottom$) limits, displaced in the same color scale. The numbers
    indicate the four selected regions used for the abundances
    determination.} \label{fig:lowhigh} \end{center} \end{figure}

\begin{figure}
  \vbox{
    \epsfig{figure=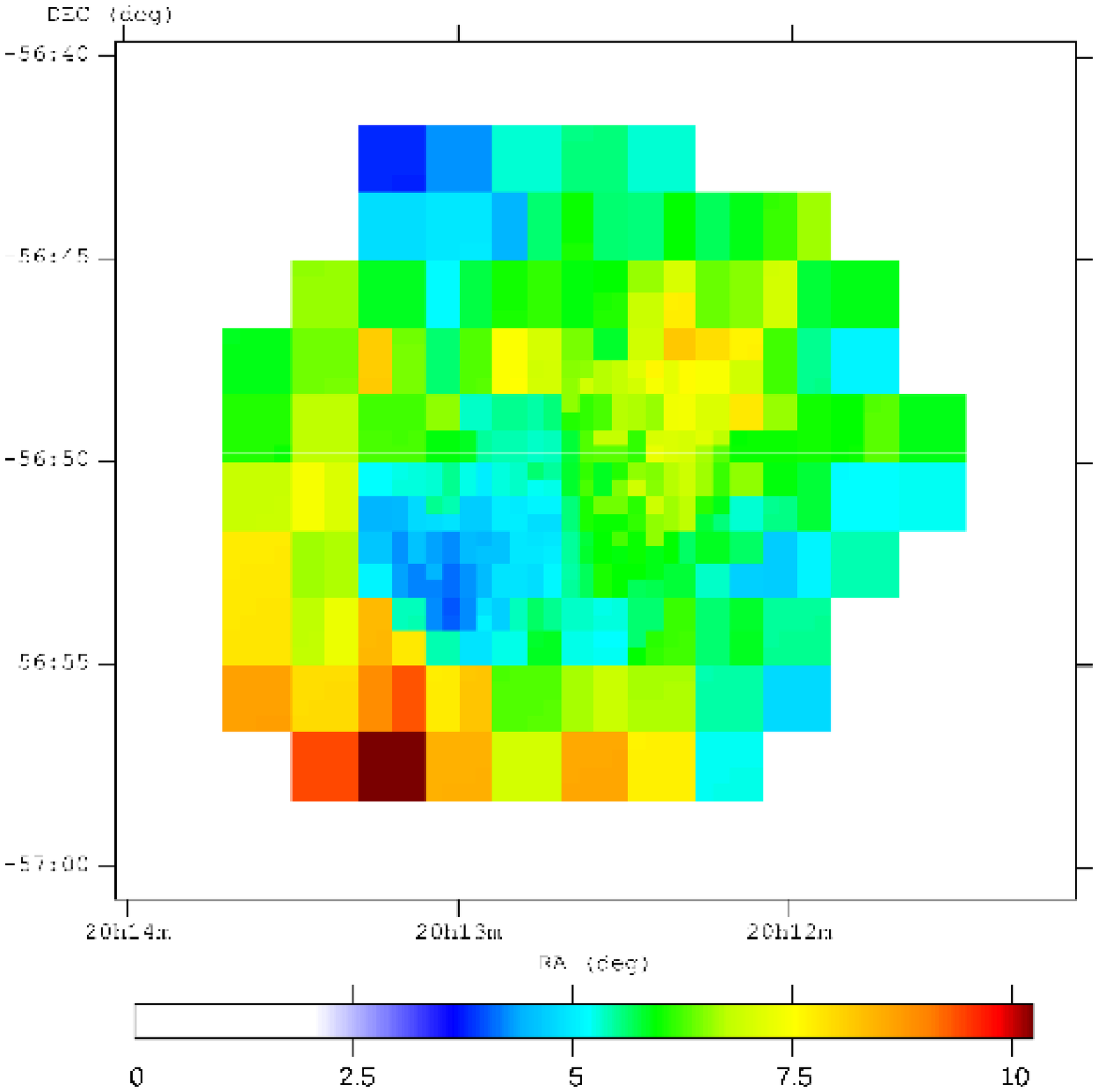,width=0.5\textwidth,height=8cm}
    \epsfig{figure=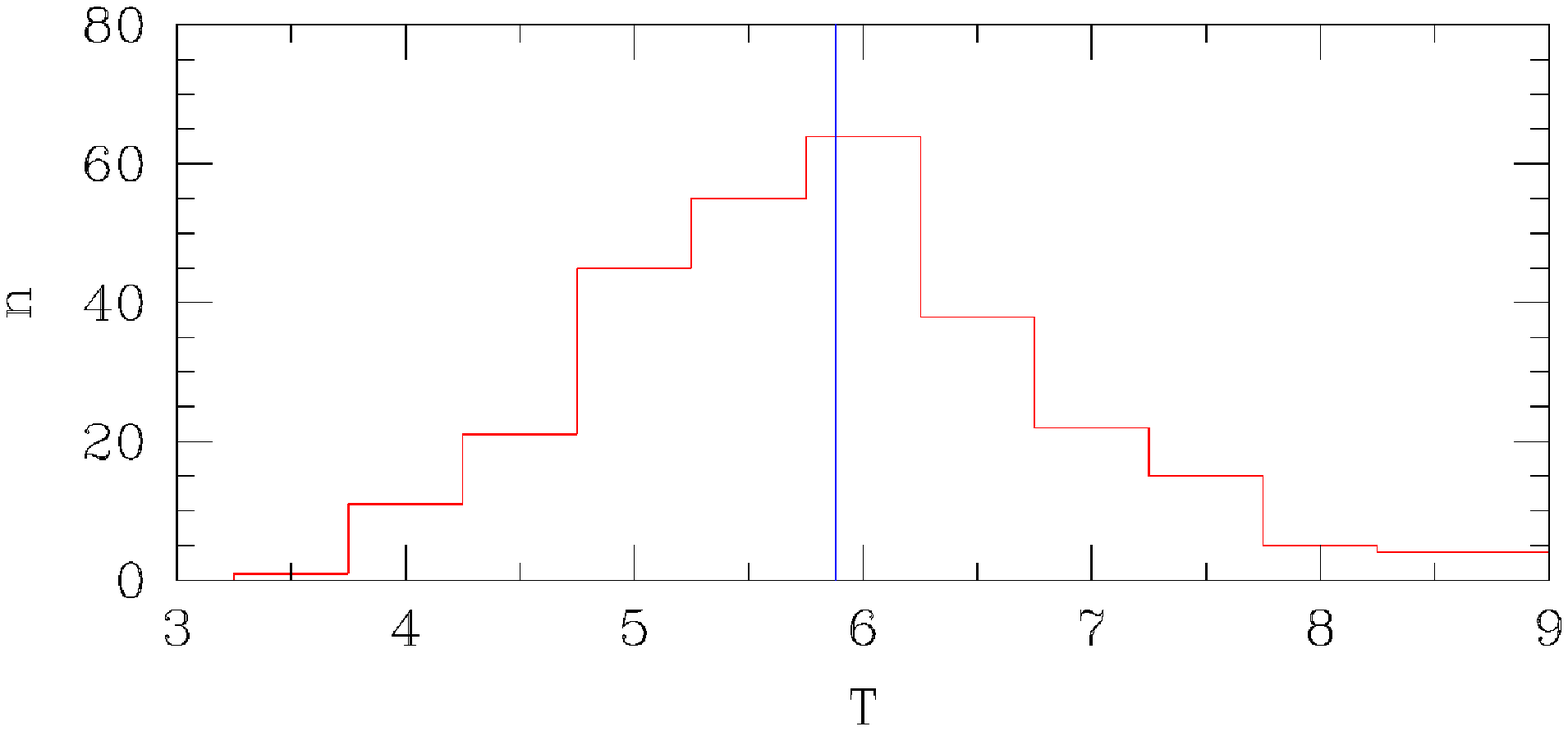,width=0.5\textwidth} } \caption{$Upper
    \ panel$: Temperature map based on spectra from all three EPIC
    cameras. $Lower \ panel$: Number of bins with a certain
    temperature. The vertical line in blue represents the mean value.
  } \label{fig:ttend.ps} \end{figure}

To obtain a metallicity measurement with a good accuracy require a
high statistic. Based on previous metallicity study we set a minimum
count number ($\sim$3,000 source counts per region) necessary for
proceeding with the spectral fit. The spectral regions for the map
were selected using the following method. We first produced an image
where each pixel is 500$\times$500 EPIC physical pixels corresponding
to 25$^{\prime \prime}\times$25$^{\prime \prime}$. So, from now on ``1
pixel'' is actually this ``fat'' 25$^{\prime \prime}\times$25$^{\prime
  \prime}$ ``pixel''. A square region with side length of $\sim$1100
arcsec and centred on the peak of the X-ray emission was defined to
include only areas where the surface brightness of the source is
high. This region was divided into 11$\times$11 square regions, each
100 arcsec$^{2}$. The size of these square regions was then optimized
by splitting it into horizontal or vertical segments through its
center, while including at least 3,000 source counts, summed over the
all three EPIC cameras. Any region which did not contain 3,000 counts
was ignored. For all the selected regions, spectra were extracted for
source and background in all three cameras. Finally, the spectra were
re-binned with the $grppha$ task, to reach at least 20 counts per
energy bin. \\ Spectra were analysed with XSPEC
\citep{1996ASPC..101...17A} version 12.3.1. Since the spectra were
re-binned, we have used standard $\chi^2$ minimization. We determined
the errors with the XSPEC tasks $error$ and $steppar$. In order to
model the emission from a single temperature we fit the spectra with
the following model:
\begin{equation}
MODEL=WABS(N_H) \times MEKAL(T,Z,K).
\end{equation}
WABS is the photoelectric absorption model by
\citet{1983ApJ...270..119M} and MeKaL model is the traditional plasma
code (\citealt{1985A&AS...62..197M}, \citeyear{1986A&AS...65..511M};
\citealt{mekal...kaastra}; \citealt{1995ApJ...438L.115L}) in which the
temperature T, the metallicity Z and the normalization K are free
parameters. The spectral fit was done leaving the hydrogen column density
to free vary. We fit jointly MOS1, MOS2 and pn spectra,
enforcing the same normalization value for MOS spectra and allowing
the pn spectrum to have a separate normalization. In the spectral
fitting we used the 0.5-8 and 0.5-7.5 keV energy range for MOS and pn
spectra respectively. We excluded the energy above 7.5 keV in the pn
spectra because of the strong fluorescence lines of Ni, Cu $\&$ Zn.
These lines, present in the background, are not well subtracted by the
double background subtraction because they do not scale perfectly with
the continuum of the particle-induced background. The redistribution
and ancillary files (RMF and ARF) were created with the SAS tasks
$rmfgen$ and $arfgen$ for each camera and each region that we
analysed. The metal abundances are based on the solar values given by
\citet{1989GeCoA..53..197A}. We chose to use these abundance for
easier comparison with previous work, although these values may not be
accurate as shown by \citet{1998SSRv...85..161G} who obtained
significantly lower O and Fe abundance then
\citet{1989GeCoA..53..197A}. In Fig. \ref{fig:57-57.ps} we show an
example of the spectra. \\ The obtained metallicity and temperature
maps are shown in Figs. \ref{fig:zzend.ps} and \ref{fig:ttend.ps}
($upper \ panel$). The resolution of the maps is the same (although it
would be possible to obtain a good estimate of the temperature within
smaller region) for a direct comparison between the two maps. Regions
in white are those where the spectral signal to noise ratio was not
sufficient to determine the metallicity. Both maps show a complex
substructure as expected for a merging cluster. In particular the
metallicity distribution appears very inhomogeneous, while the
temperature map shows a hot arc-like structure around the cold gas region.
The highest peak of the metallicity is located in the southeast with
respect to the X-ray center corresponding to the cold front region. A
higher metallicity is also observed in two blobs to the NW.
Between those clumps we note a region with a very low metallicity
(below 0.2 in solar abundances). \\ The histograms in Fig.
\ref{fig:zzend.ps} and \ref{fig:ttend.ps} ($lower \ panel$) shows the
distribution of the metallicity and temperature values. Concerning the
metallicity we note that all the values range between 0.05 and 0.7
solar abundances with a mean of $\sim$0.31 Z$_{\odot}$ that is in good
agreement with the value of 0.29 obtained by fitting the spectra of
the whole cluster within a radius of $\sim$8 arcmin centered on the
emission peak. \\ Typical errors in our metallicity map are about
10-20$\%$, although for a few pixels in the outskirts the error is
higher. In Fig. \ref{fig:lowhigh} we show the 1$\sigma$ upper and
lower limits for comparison. For the temperature maps all the errors
are lower than 10$\%$.

\begin{figure}
  \vbox{
    \epsfig{figure=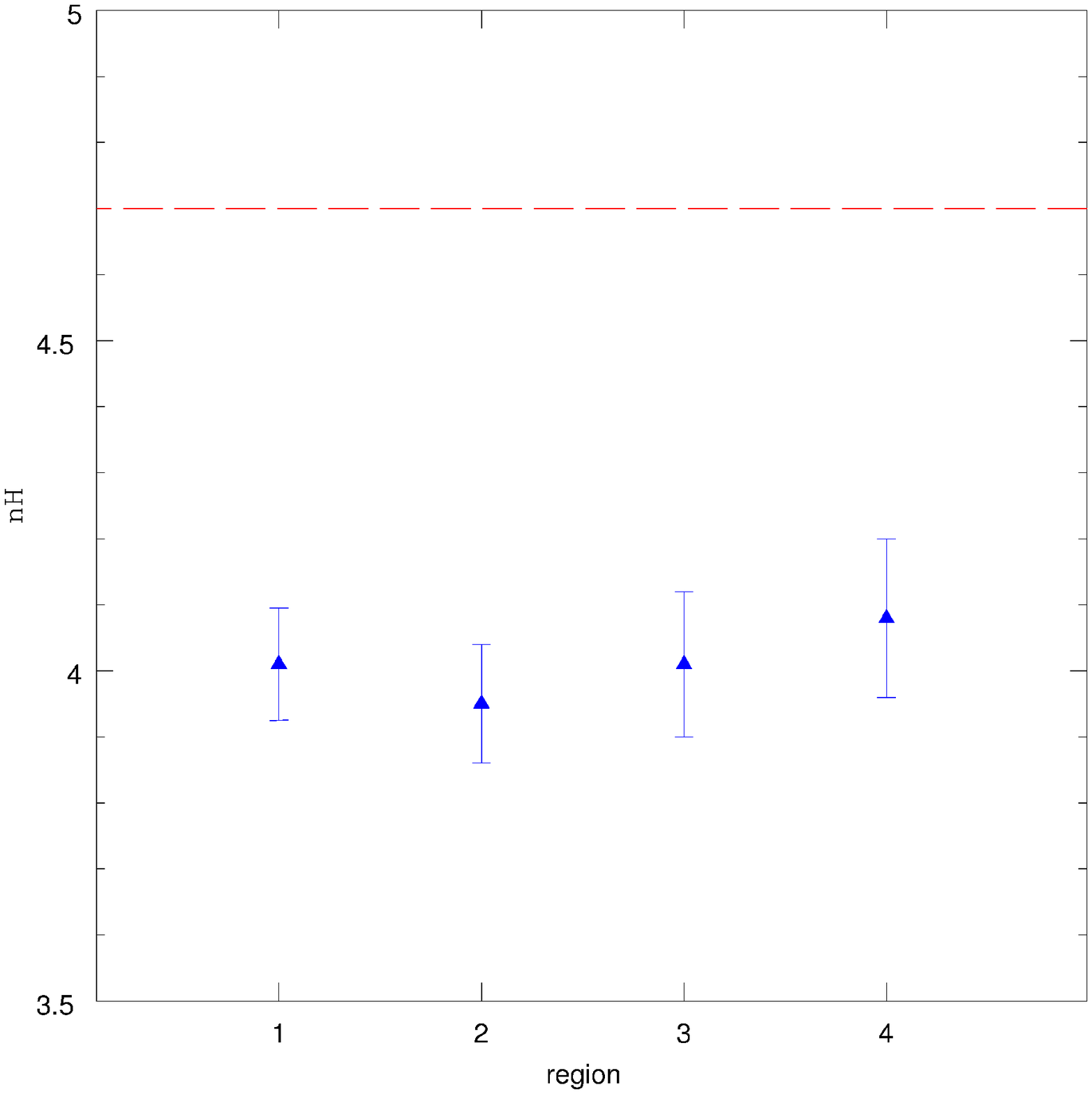,width=0.5\textwidth,height=6cm}
 \epsfig{figure=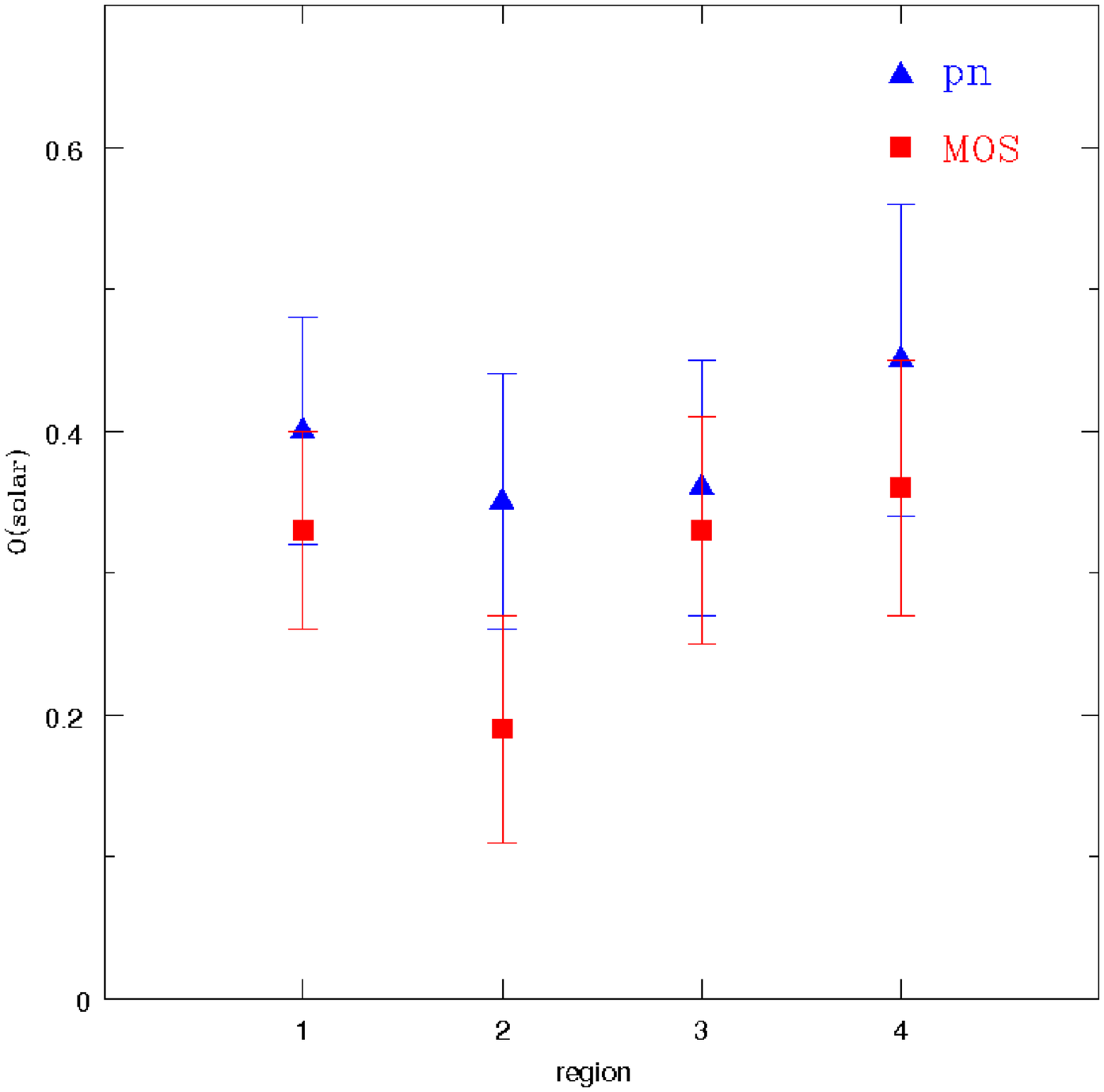,width=0.5\textwidth,height=6cm}
}
\caption{$Upper \ panel$: The hydrogen column density value obtained
  with a MeKaL model for the 4 considered regions using the all three
  EPIC cameras (MOS+pn). The red dashed line represent the value
  obtained by \cite{1990ARA&A..28..215D}. $Lower \ panel$: Oxygen
  values obtained by fitting spectra from the MOS and pn detectors
  independently with the column density fixed to the value computed
  with a MeKaL model.}
\label{fig:oxy.eps}
\end{figure}

\section{Abundances and enrichment by supernova type Ia and II}
The chemical evolution of galaxies and consequently of the ICM is
dominated by its main contributors SN Ia, SN II and planetary nebulae
(the contribution of planetary nebulae is negligible for abundances of
elements from O to Ni). We investigated the relative contribution of
the supernova type Ia/II to the total enrichment on the intra-cluster
medium. We assumed that the total number of atoms N$_{i}$ of the
elements $i$ is a linear combination of the number of atoms Y$_{i}$
produced per supernova type Ia (Y$_{i,Ia}$) and type II (Y$_{i,II}$):

\begin{center}
\begin{equation}\label{eq:supernovae}
N_{i}=\alpha Y_{i,Ia} + \beta Y_{i,II},
\end{equation}
\end{center}
where $\alpha$ and $\beta$ are the numbers of supernova types Ia and
II respectively. We used SNIa yields obtained from two different models
adapted from \citet{1999ApJS..125..439I}: the W7 model is a so-called
slow deflagration model, while WDD2 is the favored model by
\citet{1999ApJS..125..439I} and is calculated using a delayed
detonation model. For nucleosynthesis products of SN II we adopted
average yields of stars in a mass range from 10 M$_{\odot}$ to 50
M$_{\odot}$ calculated by \citet{1995MNRAS.277..945T} assuming a
Salpeter initial mass function (IMF). \\ From observations we obtained
the number of nuclei per hydrogen nucleus relative to the solar
abundances:
\begin{equation}\label{eq:meka}
\frac{N_i}{N_H}=f_{i}{\left( \frac{N_i}{N_H} \right)}_{\odot},
\end{equation}
where N$_H$ is the number of hydrogen atoms and $f_i$ the abundance
obtained directly from XSPEC analysis.  Combining equations
\ref{eq:supernovae} and \ref{eq:meka} and considering iron as
fixed element we can compute the ratio between supernova type Ia and
II:
\begin{center}
\begin{equation}
\frac{\alpha}{\beta}=\frac {{\left( \frac{Fe}{f_{Fe}} -
    \frac{N_i}{f_i} \right)}_{SN II}} {{\left( \frac{N_{i}}{f_{i}} -
    \frac{Fe}{f_{Fe}} \right)}_{SN Ia}},
\end{equation}
\end{center}
where $Fe$ in the equation is the yield value for the iron. We chose
to fix iron because is the best determined element. \\ We wanted to
determine if the high metallicity region is associated with a
particular kind of SN type or with high number of SNe II as
consequence of an intense episode of star formation, due for instance
to ram-pressure effect \citep{2009A&A...499...87K}. To do this we
selected four regions: first we extracted the spectra from a circular
region, centered on the X-ray peak, with a radius of 3$^{\prime}$
(region 2 in $lower \ panel$ of Fig. \ref{fig:lowhigh}) ; then we
selected an annulus with inner and outer radius of 3$^{\prime}$ and
7$^{\prime}$ respectively and we extracted a spectrum for the low
metallicity region by selecting a sector between PA = 180$^{\circ}$
and 90$^{\circ}$ (East to West) and for the high metallicity region
selecting the sector between PA = 90$^{\circ}$ and 180$^{\circ}$
(respectively region 3 and 4); finally, we extracted the spectra from
a circle with a radius of 7$^{\prime}$ (region 1) that includes all
the three previous regions.

\subsection{Abundances determination}
In Table \ref{table:abb} we show the obtained abundances with their
$\pm$1$\sigma$ errors for one parameter for the four selected
regions. We fitted the data with the following procedure to avoid the
degeneracy of the parameters: (1) we fitted the data with an absorbed
MEKAL model in the 0.4-7 keV band to obtain temperature and nH
(metallicity and normalization are considered free parameters); (2) we
fixed nH and temperature and use a VMEKAL model in the same energy
band to determine the iron abundance (O, Si, S, Ar are left free, the
other elements are fixed to the solar value); (3) we kept temperature
and iron fixed to measure oxygen abundance in the 0.4-1.5 keV band;
(4) we fix the values of temperature, iron and oxygen to estimate the
silicon, sulfur and argon abundances in the 1.5-5 keV band. \\ We
fitted the element abundances in narrow bands around to the
corresponding emission lines, allowing the normalization to vary, in
order to correct for small inaccuracies in the best determination of
the continuum in those narrow energy band. \\ The Ne and Mg abundances
could not be constrained because these lines are blended with the Fe-L
complex at the EPIC spectral resolution. The aluminium line is blended
with the much stronger silicon line and is not measurable. The Ni
abundance determinations are driven almost entirely by the He-like and
H-like K-shell lines at 7.77 and 8.10 keV which are both beyond the
chosen spectral fitting band. \\ In Fig. \ref{fig:oxy.eps} ($upper
\ panel$) we show the hydrogen column density obtained in all the four
analyzed regions. We note that is always lower than the Galactic value
of 4.7$\times$10$^{20}$ cm$^{-2}$ determined from the 21 cm radio
observation \cite{1990ARA&A..28..215D}. We left free nH (not fixed to
the Galactic value) because of the large discrepancy and also because
the O abundance determination is sensitive to the presence of excess
absorption and to the cross-calibration uncertainties between the
spectral response of the two EPIC instruments in the soft energy band
(below 1 keV). In the $lower \ panel$ of Fig. \ref{fig:oxy.eps} we
show the oxygen abundance in the four different regions obtained by
fitting the spectra from the MOS and pn detectors separately with
N$_H$ fixed at the value obtained with a MEKAL model in the 0.4-7 keV
band. The agreement between the two detectors is quite good; the
discrepancy in the central 3$^{\prime}$ (region 2) can be due to a
calibration problem at this particular position of the detector.

\setlength{\extrarowheight}{0.1cm}
\begin{table}
\caption{Abundances obtained by fitting the four selected regions
  shown in Fig. \ref{fig:lowhigh}. Temperatures are given in keV,
  N$_H$ in 10$^{20}$ cm$^{-2}$ and abundances are given with respect
  to solar. All errors quoted are at the 68$\%$ level for one interesting parameter ($\Delta {\chi}^2$=1).}
\label{table:abb}
\centering
\begin{tabular}{l | c c c c}
\hline\hline
Par. & 1 & 2 & 3 & 4 \\
\hline
N$_H$ & 4.01$\pm$0.08 & 3.95$\pm$0.09 & 4.01$\pm$0.11 & 4.08$\pm$0.12 \\
kT    & 5.919$\pm$0.035 & 6.157$\pm$0.053 & 6.516$\pm$0.068 & 5.050$\pm$0.049 \\
O     & 0.34$\pm$0.05 & 0.26$\pm$0.06 & 0.30$\pm$0.07 & 0.40$\pm$0.07 \\
Si    & 0.43$\pm$0.07 & 0.37$\pm$0.08 & 0.35$\pm$0.07 & 0.52$\pm$0.08 \\
S     & 0.18$\pm$0.10 & 0.21$\pm$0.10 & 0.12$\pm$0.10 & 0.22$\pm$0.10 \\
Ar    & 0.28$\pm$0.11 & 0.27$\pm$0.18 & 0.24$\pm$0.20 & 0.36$\pm$0.20 \\
Fe    & 0.306$\pm$0.009 & 0.288$\pm$0.010 & 0.275$\pm$0.014 & 0.369$\pm$0.013 \\
\hline
$\chi^{2}_{Red}$ & 1.58 & 1.19 & 1.23 & 1.23 \\
\hline
\end{tabular}
\end{table}

\setlength{\extrarowheight}{0.1cm}
\begin{table}
\caption{Relative number of SN II contributing to the enrichment of
  the intra-cluster medium. We show the results for different region
  extractions using the yields of SN Ia and SN II, with two physically
  different SN Ia yield models. In the second column we indicate the
  element used in combination with iron to determine the percentage number of SN
  II.}
\label{table:SN}
\centering
\begin{tabular}{l | c c c c c}
  \hline\hline
  Model & el. & 1 & 2 & 3 & 4 \\
  \hline
  W7     & O  & 72.6$^{+3.7}_{-4.6}$ & 66.5$^{+6.1}_{-7.9}$ & 72.0$^{+5.8}_{-6.7}$ & 71.9$^{+4.4}_{-5.7}$  \\
  & Si & 75.0$^{+6.6}_{-9.9}$ & 70.3$^{+9.5}_{-16}$ & 69.8$^{+8}_{-15}$ & 75.1$^{+6.3}_{-9.2}$  \\
  & S  & $<$55.4   & $<$70.4   & $<$36.6   & $<$48.8    \\
  & Ar & 83$^{+12}_{-45}$ & 84$^{+15}_{-84}$ & 81$^{+19}_{-81}$ & 86$^{+23}_{-86}$  \\
  \hline
  WDD2   & O  & 75.5$^{+3.4}_{-4.2}$  & 70.0$^{+5.5}_{-7.4}$  & 75.0$^{+5.2}_{-7.1}$  & 74.9$^{+3}_{-5.1}$    \\
  & Si & 74.3$^{+7.4}_{-11.8}$  & 69.0$^{+11}_{-21}$ & 68.3$^{+10.4}_{-19}$ & 74.5$^{+7.0}_{-11}$    \\
  & S  & $<$23.8       & $<$59.5       & -             & $<$2.1        \\
  & Ar & 75$^{+19}_{-75}$ & 77$^{+22}_{-77}$ & 70$^{+30}_{-70}$ & 81$^{+18}_{-81}$ \\
  \hline
\end{tabular}
\end{table}

\begin{figure}
  \epsfig{figure=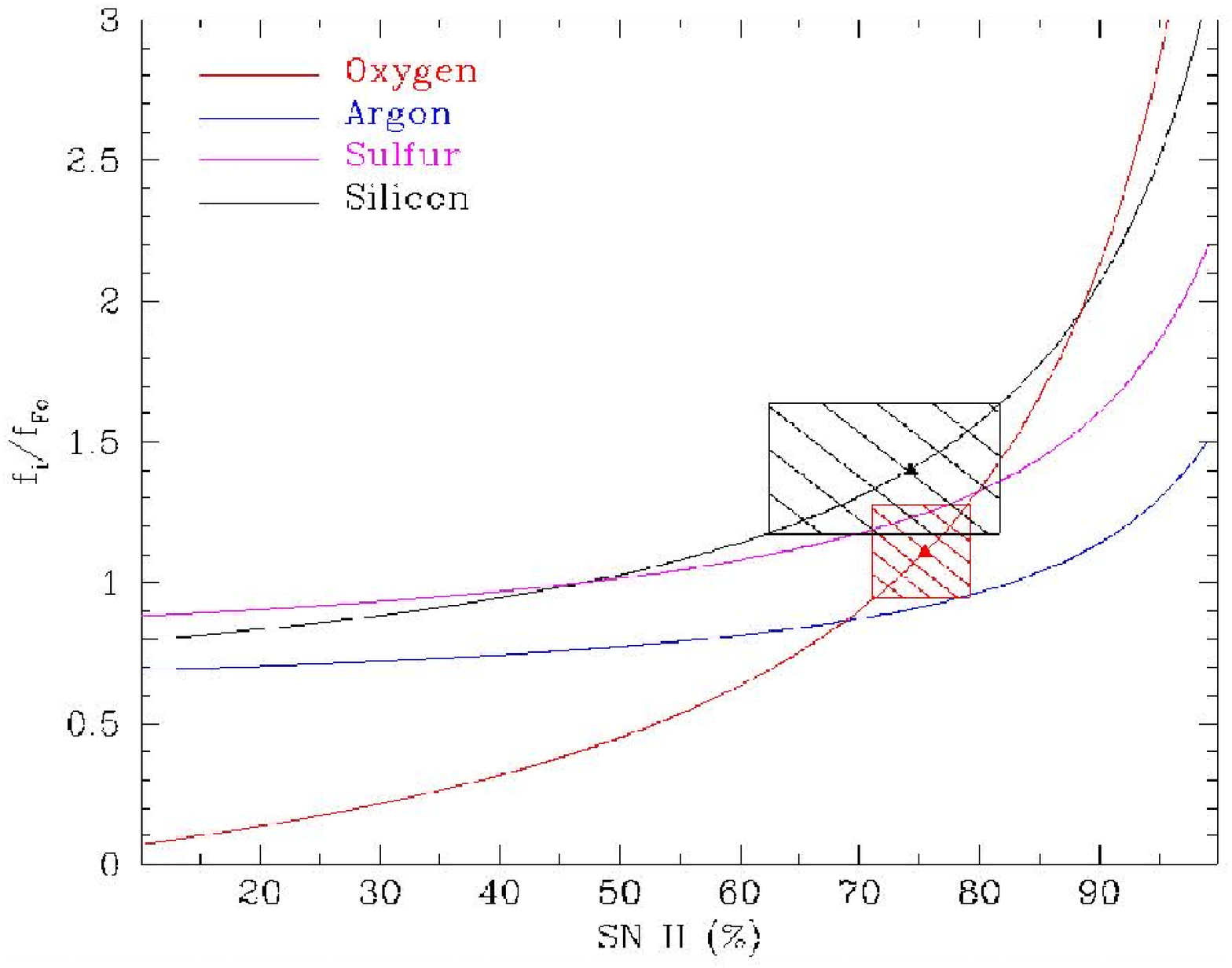,width=0.5\textwidth,height=7.5cm}
  \caption{The lines represent the abundance ratios as the SN ratio
    varying for the different elements. The two solid triangles
    present the observed O/Fe (red) and Si/Fe (black) ratio for the
    7$^{\prime}$ region while the boxes present the error associated
    with the observed abundance (Y axis) and the relative number of SN
    II associated with the lower and upper limit (X axis). }
  \label{fig:theorySN} \end{figure}

\section{Discussion}
\subsection{Metallicity map}
There are a number of features to note analyzing the metal
map. \\ First, the metal distribution is very inhomogeneous with
several maxima and complex metal patterns as expected for a merging
cluster \citep{2006A&A...447..827K}. The X-ray peak is located in a
region with a very low metallicity (0.15-0.20 Z$_{\odot}$, see Fig.
\ref{fig:zzend.ps}) with respect to the mean metallicity of the cluster
(0.31 Z$_{\odot}$). From the simulations is clear that the maximum of
the metallicity is not always in the cluster center
\citep{2006A&A...447..827K}.  The reason is, that enriched gas, that
might fall into the center, is mixed with a lot of other gas as in the
centre the gas density is high.  Therefore it hardly increases the
metallicity there. If, however, a starburst happens in the outer parts
of the cluster, the enriched gas mixed only with a small amount of
other gas and therefore it can increase the metallicity there
considerably. Hence it can happen that the maximum of the metallicity
is temporarily not in the cluster centre. \\ Althouh several blobs
with high metallicity ($\sim$0.5 Z$_{\odot}$) are present in the NW
direction along the axis of the X-ray elongation we found that the
most metal rich zone correspond to the cold front region, as
previously shown by \citet{2004A&A...426....1B}, with a peak of 0.7
solar abundance.  This region probably shows the most interesting
feature in the metal map and it can be used together with simulations
to infer the dynamical state of the cluster.

\subsection{SN enrichment}
In Table \ref{table:SN} we show the obtained best fit values of the
relative contribution of SN II with a confidence level of 68$\%$. We
see that both models are consistent with a scenario where the relative
number of supernovae type II contributing to the enrichment of the
intra-cluster medium is 55-95$\%$ depending on the considered elements
and regions. The best agreement between the data of O, Si, Ar and Fe
is obtained using the WDD2 model, although the error bars are quite
large due to the uncertainties both in the observations and in the
theoretical yields.  The relative number of SN II seems to be higher
in the metallicity peak region (4), and lower for the regions 2 and
3. The lowest value is obtained in the center (region 2) and it is
consistent with the idea of an excess of SN Ia in the cD galaxies
(\citealt{2008SSRv..134..337W}).\\ The measured abundances of S are
quite low and also considering the 1$\sigma$ upper limit the
percentage of SN II that we derive is not consistent with the results
given by the other elements. We note that the relative low value of
0.18 obtained for the abundances of S in a radius of 7$^{\prime}$ is
in good agreement with the value of 0.20 computed by
\citet{2004A&A...426....1B} for the whole cluster. This value agrees
also with the result of the sulfur abundance obtained with ASCA data
for a sample of clusters with the same temperature as A3667
\citep{2005ApJ...620..680B} confirming both the prevalence of SN II in
the enrichment and a reduced S yield in the SN II model.\\ We note
that for all the elements the lower relative error is larger than the
upper one in the relative SN II determination. To explain the reason
for it, we combine the equations \ref{eq:supernovae} and
\ref{eq:meka}, and we find
\begin{equation}
\frac{f_i}{f_{Fe}}=\frac{N_{i,II}\frac{\alpha}{\beta}+N_{i,I}}{Fe_{II}+Fe_I\frac{\alpha}{\beta}},
\end{equation}
that gives the theoretical abundance ratio $f_i/f_{Fe}$ for varying SN
type Ia/II ratios. We show in Fig. \ref{fig:theorySN} the $f_i/f_{Fe}$
obtained using the yields of the WDD2 model for supernovae type Ia. It
is clear that not all the values of $f_i/f_{Fe}$ are allowed: for
example the Ar/Fe abundance ratio must be in the range between 0.7 and
1.5. In Fig. \ref{fig:theorySN} we see that all the curves grow very
slowly for a low percentage of SN II and they become very steep when
that percentage increases. This implies that the determination of the
SN fraction becomes quite inaccurate when the relative number of SN II
is low, in particular for S/Fe and Ar/Fe that shows a very flat curve
when the number of SN II is lower than 60$\%$ of the total supernovae.
If we consider symmetric error bars for $f_i/f_{Fe}$ it is clear why
for the lower relative error in the SN II determination is always
larger than the upper one. This does not influence too much the
determination of the SN relative number only if the curve is very
steep as for example the case of the O/Fe ratio. \\ To obtain the best
agreement between the observed abundances of O, Si, Ar and Fe for the
7$^{\prime}$ region we estimated that 65-80$\%$ of SN II are
necessary. Other authors tried to determine the ratio of SN Ia to SN
II events in relaxed galaxy clusters by aiming for a best fit to an
overall solar abundance pattern from O to Ni.
\citet{2006A&A...449..475W} found that the number contribution of SN
II with respect to the total number of supernovae is $\sim$75$\%$,
\citet{2006A&A...452..397D} constrained this number to the range
50-75$\%$ while \citet{2009A&A...493..409S} estimated a 30-40$\%$
contribution by SN Ia compared to type II. These results suggest that
so far the enrichment of the ICM is mainly due to SN II. We note that
\citet{2002NewA....7..227P} using different chemical evolution models
for galaxies, showed that SN II dominate the chemical enrichment
inside the galaxies, while Ia supernovae play a predominant role in
the ICM, that is not in agreement with the observational
results. \\ When we interpret the supernovae ratio we have to take
into account that the abundances ratio does not only depend on stellar
yields and IMF but also on the timescales of production of various
elements \citep{2005PASA...22...49M}. The abundance ratios will tend
to the ratios of their yield per stellar generation only if the global
metal production is considered (metals in stars, gas inside and
outside the galaxies), but it fails if only the metals in the
individual component are taken into account (e.g. the gas of
ICM). Thus, the supernovae estimation listed above should be
interpreted as the number of supernovae that would be needed to
reproduce the same abundances observed in the ICM, and not the number
of supernovae during the history of the cluster. \\ It is interesting
to compare the estimated values with the ones obtained for the
galaxies.  \citet{2008PhDT........16L} shows the results of the Lick
Observatory Supernova Search (LOSS) and in particular he focuses on
the determination of the supernovae in the local universe. He found
that about 62$\%$ of the SNe observed in galaxies are SN II. In order
to reproduce the observed abundances, \citet{1995MNRAS.277..945T}
determined the percentage of SN II for our Galaxy to be $\sim$87$\%$,
while it is $\sim$77$\%$ and $\sim$83$\%$ for the Large and Small
Magellanic Clouds respectively. The relative contribution to the
enrichment of ICM by SN II in A3667 is between these values.

\begin{figure*}
  \vbox{
    \epsfig{figure=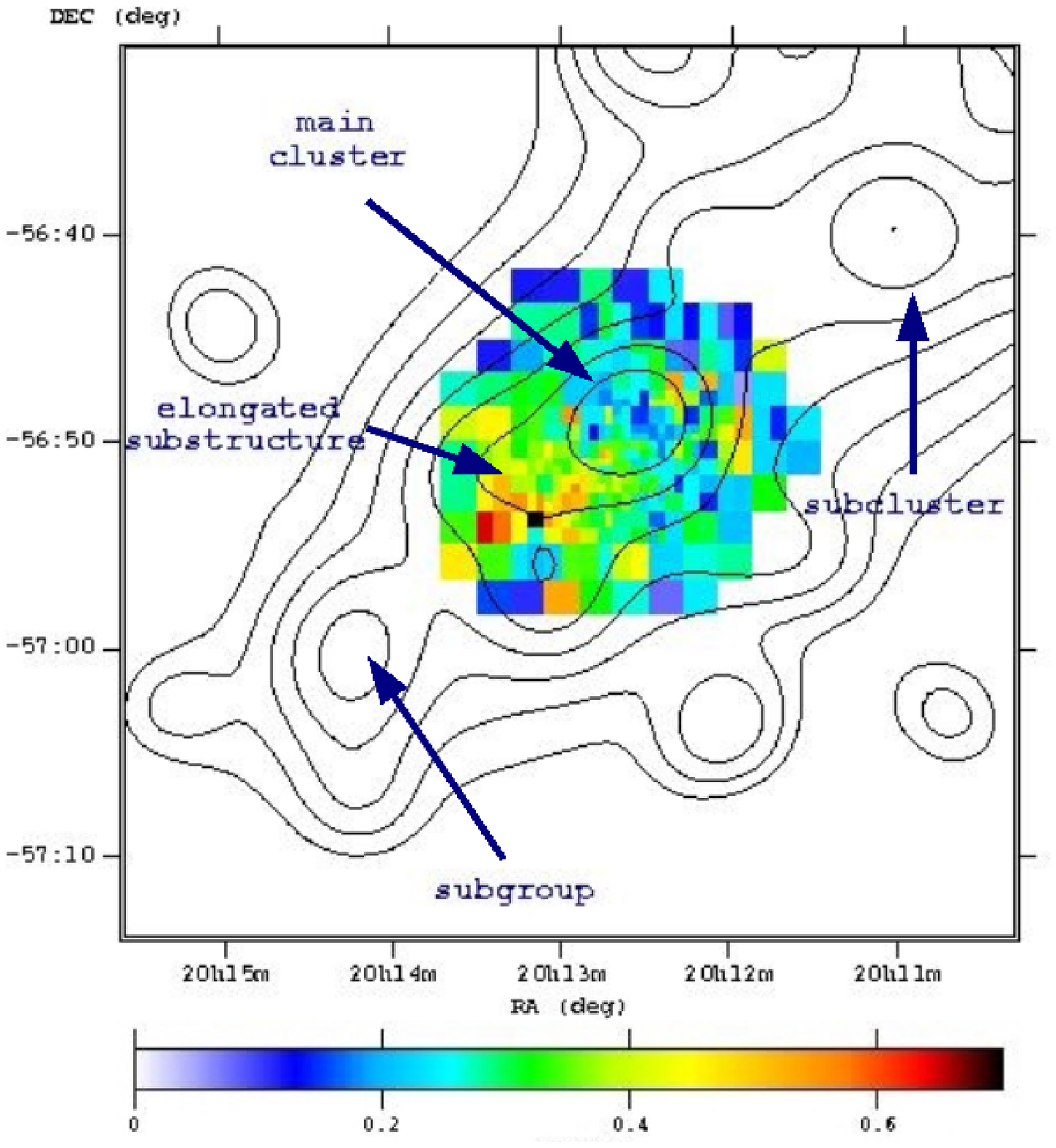,width=0.5\textwidth,height=8.5cm}
 \epsfig{figure=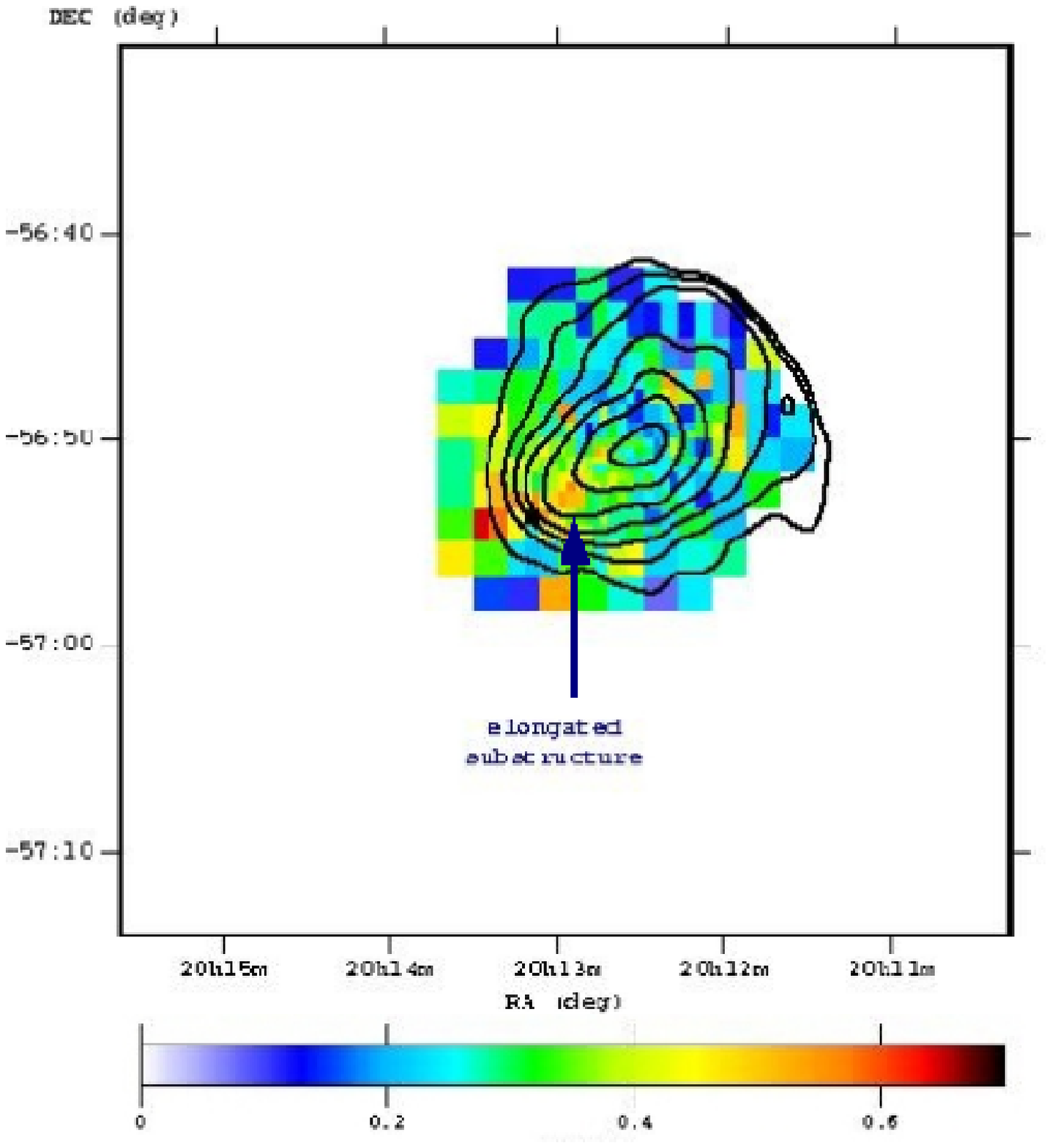,width=0.5\textwidth,height=8.5cm}
} \caption{Metallicity maps with the galaxies isodensity ($left$)
and X-ray ($right$) contours superimposed. The scale of the
metallicity is in solar units. An elongated substructure, both in
optical and X-ray contours is shown.} \label{fig:zzcon.ps}
\end{figure*}

\subsection{Comparison to simulations}
\begin{figure}
  \epsfig{figure=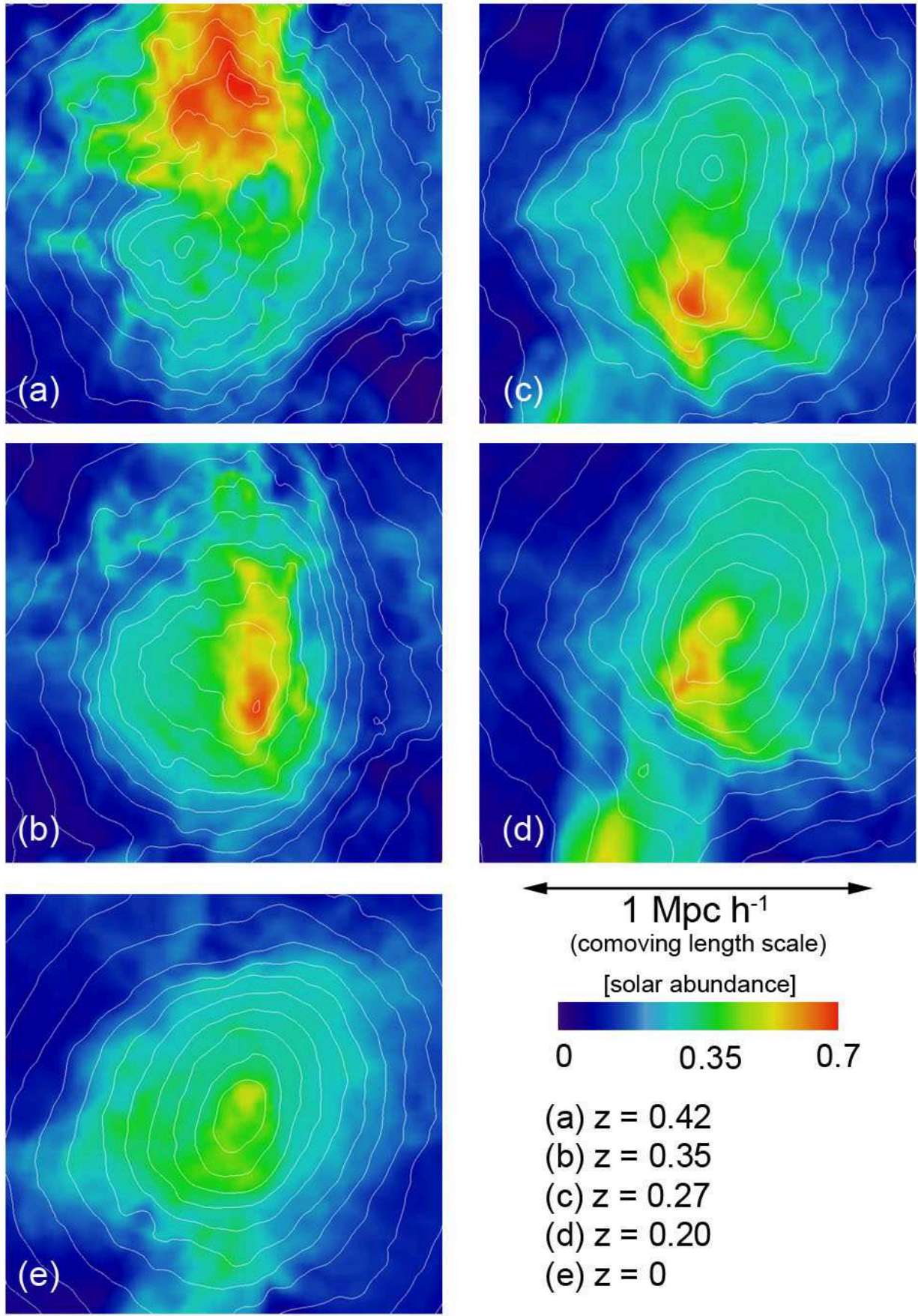,width=1\textwidth} \caption{Simulated X-ray
    weighted metal maps. X-ray surface brightness contours are
    overlaid. A model cluster showing similar features in the
    metallicity distribution as A3667 as been selected. The five maps
    correspond to different redshifts: (a) z=0.42, (b) z=0.35, (c)
    z=0.27, (d) z=0.2 and (e) z=0.} \label{sim} \end{figure}

Simulations of the enrichment of the ICM are a powerful tool to
investigate the strengths of different enrichment processes and their
spatial and temporal behavior. To study the origin of the
inhomogeneities found in the metal distributions in many galaxy
clusters, several simulations were performed. In the numerical setup
we used different code modules to calculate the main components of a
galaxy cluster in the framework of a standard $\Lambda$CDM
cosmology. The non-baryonic dark matter (DM) component is calculated
using GADGET2 \citep{2005MNRAS.364.1105S} with constrained random
field initial conditions \citep{1991ApJ...380L...5H}, implemented by
\citet{1996MNRAS.281...84V}. For the treatment of the ICM we use
comoving Eulerian hydrodynamic with a shock capturing schemes (PPM,
\citealt{1984JCoPh..54..174C}), with a fixed mesh refinement
\citep{1992A&A...265...82R} on four levels and radiative cooling
\citep{1993ApJS...88..253S}. The properties of the galaxies are
calculated by an improved version \citep{2005MNRAS.359..469V} of the
galaxy formation code of \citet{1999MNRAS.310...43V} which is a
semi-analytic model in the sense that the merging history of galaxy
halo is taken directly from the cosmological N-body simulation. With
this setup we investigated two different enrichment processes, namely
supernova driven galactic winds and ram-pressure stripping (see
\citealt{kap...2009} for details regarding the galactic winds model
and \citealt{2006A&A...452..795D} for the ram-pressure stripping
model).\\ In Fig. \ref{sim} the evolution of the metallicity in a
model cluster is shown. The isolines correspond to the X-ray surface
brightness of the model cluster. In the simulation a region of
enriched material falls towards the cluster center. The high
metallicity is in this infall region (see at the top of panel (a) in
Fig. \ref{sim}) and it is caused by four starbursts (with an outflow
rate of more than 25 solar masses each) that happened before z=0.42.
Approaching the cluster center the material gets mixed with the lower
metallicity ICM along the trajectory (see panel (b) in
Fig. \ref{sim}). From the turnaround point (see panel (c)
Fig. \ref{sim}) the material falls smoothly to the center and gets
mixed by the ambient ICM with lower metallicity (see panel (d)
Fig. \ref{sim}). As the material ejected by galactic winds and
starbursts contains more SNII products the feature in the simulation
(panel (d) in Fig. \ref{sim}) corresponds nicely to the off-center metal
concentration found in A3667 (see Fig. \ref{fig:zzend.ps}) where the
relative contribution of the SN II is higher (region 4). At a redshift
of z=0 the enriched gas originating from starbursts has already mixed
with the ICM, leading to the metal map shown in panel (d) in Fig.
\ref{sim}. \\ Metal blobs originating either from galactic winds or
ram-pressure stripping are a common feature in metal enrichment
simulations. Typically they move along the trajectory of the
underlying galaxies and as they start to feel the pressure of the
surrounding gas they lag behind the originating galaxies and mix with
the surrounding gas. The more inhomogeneities are present in the ICM
the more recent the enrichment processes took place. Therefore the
inhomogeneities found in the metal maps in galaxy clusters are
indicators for the merging frequency of substructures with the
cluster. Typically the inhomogeneities in the metal maps vanish over
timescales of several 100 Myrs to several Gyrs depending to the mass
present in the metal feature. In the example in Fig. \ref{sim} the
metallicity blob survives nearly 3 Gyrs. \\ Based on the size of the
high metallicity region ($\sim$3$^{\prime}$ of radius), we expect that
the metal feature in Fig. \ref{fig:zzend.ps} is either a consequence
of a recent merger or of an older merger but which involve a larger
mass. In the latter case the metal feature would have survived for a
long time.

\subsection{Dynamical state}
We produced a multi-scale galaxy density map (see
  \citealt{2005A&A...430...19F} for more details) using the 550
  spectroscopically confirmed cluster members obtained by
  \citet{2009ApJ...693..901O}. In Fig. \ref{fig:zzcon.ps} we plotted
  for comparison the metal map with the galaxy isodensity overlaid
  ($left$) and X-ray ($right$) contours. As for the X-ray surface
  brightness the projected galaxy density map shows an elongation in
  the direction of the two radio relics.  The comparison of the ICM
  metallicity distribution and the position of the sub clusters of
  A3667 can give hints on the complex merging scenario of this cluster
  \citep{2006A&A...447..827K}. We detected a metal peak between the
  main cluster and the SE subgroup. According to
  \citet{2006A&A...447..827K} we expect high metallicity between
  subclusters in a post-merger phase. Thus, this configuration
  supports the scenario suggested by \citet{2009ApJ...693..901O} in
  which the SE subgroup has traveled from the NW and passed through
  the main cluster where the ram pressure stripped off the enriched
  and cooler gas. \\ However, an elongation towards the high
  metallicity peak visible both in optical and X-ray images (see the
  contours in Fig. \ref{fig:zzcon.ps}) could suggest a more complex
  dynamics in the cluster center.  The abundances of the measured
  elements are higher in region 4 with respect to the regions 2 and
  3. This result can be explained if a group of galaxies, located in
  the elongated substructure and containing both SN Ia and SN II
  products, falls into a cluster moving from southeast to
  northwest. The inter-stellar medium is thus stripped off by
  ram-pressure stripping and leads to a more peaked abundance
  distribution. On the other hand, in region 4 the relative number of
  SN II seems to be higher, with respect to the other two considered
  regions (region 2 and 3), suggesting that the metallicity peak in
  region 4 is mainly due to galactic winds as obtained in the
  simulations. Due to the large error bars in the SN determination the
  latter result has to be confirmed with a deeper
  observation. \\ Interestingly a gap in metallicity has been detected
  by \citealt{2004A&A...426....1B} in between the main cluster and the
  NW sub-cluster (i.e. a region not covered by our metallicity
  map). Based on the results of \citep{2006A&A...447..827K} this would
  imply that these two structures are in a pre-merger phase. All these
  results could suggest that A3667 is a cluster forming through
  multiple merging events along a common NW-SE axis.

 \section{Summary}
We analyzed a 64 ks \xmm exposure of the merging cluster of galaxies
A3667. We obtained a detailed 2D metallicity map. From this we can
conclude that:
\begin{itemize}
\item the distribution of metals is clearly non-spherical. It looks
  very inhomogeneous with several maxima separated by very low
  metallicity regions; \item the highest metallicity peak is located
  on the southeast with respect to the X-ray center and it corresponds
  to the region with the lowest temperature.
\end{itemize}
We also measured the abundances for oxygen, silicon, sulfur, argon and
iron in 4 different regions of the cluster and determined the
number ratio of supernovae type Ia and type II. From these data we
conclude that:
\begin{itemize}
\item using the elements abundance of Fe, O and Si we found that the
  relative number of supernovae type II necessary to reproduce the
  observed abundances in A3667 ranges between 65-80$\%$;
 \item the delayed detonation model WDD2 seems to reproduce the
   observed data better compared to the slow deflagration model; \item
   the supernovae number estimation from the abundances of sulfur is
   not in agreement with the estimate obtained using the other
   elements confirming a reduced S yield in the SN II model.
\end{itemize}
\noindent Finally, we discussed the dynamical state of the
  cluster by comparing the ICM metal and galaxy density maps to our
  simulations. In agreement with the scenario proposed by
  \citet{2009ApJ...693..901O}, we conclude that the SE subgroup moved
  from the NW and passed through the main cluster, where the ram
  pressure stripped off the enriched and cooler gas as seen in the
  metallicity and temperature maps. The highest metallicity region,
  that shows a higher contribution of SN II, could be partly related
  to an enrichment by galactig winds due to star formation possibly
  triggered by an infalling group. In addition, two metal rich blobs
  in the NW of the main cluster could partly result from
  inhomogeneities not completed dispersed after an old merger, which
  is possibly responsible for the formation of the two radio
  relics. Based on the comparison with previous X-ray data
  (e.g. \citealt{2004A&A...426....1B}) we conclude that A3667 has a
  complex dynamical history and it is possibly evolving by accreting
  sub-clusters along a main NW-SE axis.

\vspace*{10pt}
\noindent
\small{\emph{Acknowledgements.} We warmly thank Matt Owers for
  providing the catalog with positions of confirmed cluster members
  and Jean-Patrick Henry the referee for very useful comments.  The
  authors acknowledge the Austrian Science Foundation (FWF) through
  grants P18523-N16 and P19300-N16.}

\bibliographystyle{aa} \bibliography{12933}

\end{document}